\documentstyle[11pt]{article}

\begin{document}
\begin{center}
\begin{flushright}
{BIEHP-TH-96-14\\
hep-ph/9604342}
\end{flushright}

\vspace{10mm}
{\Large Strange quark distribution and corrections due to
shadowing and isospin symmetry breaking}

\vspace{10mm}
\renewcommand{\thefootnote}{\fnsymbol{footnote}}
{\normalsize Ma Boqiang}

\vspace{8mm}
{\large
Institute of High Energy Physics, Academia Sinica, P.O.Box 918(4),
Beijing 100039}
\vspace{4mm}

{\large \bf Abstract } 
\end{center}

The strange sea quarks distributions of nucleons
obtained by two global
analyses based on available structure function data of muon and
neutrino deep inelastic scatterings
are different from the strange sea quark distribution
measured by the  CCFR Collaboration
from dimuon events in neutrino scattering.
We discuss possible contributions to this
discrepancy 
from the
nuclear shadowing in the deuteron and from
the isospin symmetry breaking in the sea between the
neutron and the proton.

\vspace{12mm}

PACS number(s): 14.20.Dh, 13.85.Qk, 13.60.Hb, 24.85.+p

\vspace{4mm}
Published in Chinese Physics Letters {\bf 13} (1996) 648-651.
\break

The global analyses of quark distributions of nucleons
are undergoing rapid progress due to the increased
precision of deep inelastic lepton scattering data \cite{OT92}.
The New Muon Collaboration (NMC) data
on $F_{2}^{p,d}$ from muon scattering \cite{NMC92} and
Columbia-Chicago-Fermilab-Rochester (CCFR) data
on $F_{2,3}^{Fe}$ from neutrino and antineutrino
scatterings \cite{CCFR92} have been used by the
CTEQ Collaboration \cite{CTEQ93} and the Durham-RAL
group (MRS) \cite{MRS93} in their global analyses
of quark distributions.
The new strange sea quark distributions are found to be
larger than those of earlier fits.
A strange quark distribution from a
leading-order QCD analysis of opposite-sign dimuon events
induced by neutrino scattering has been presented
by the CCFR Collaboration recently \cite{CCFR93}.
From Fig.~1, where the CCFR, CTEQ and MRS results of strange
quark distribution are presented, we see that there is
a discrepancy between the strange quark distribution
from dimuon events in neutrino scattering and those
of the global analyses.
We will discuss in this paper the contributions to this discrepancy
from the
nuclear shadowing effect in the deuteron and from
the isospin symmetry breaking in the sea between the
neutron and proton.

The CTEQ and MRS global analyses are mainly based on data of
structure functions from
muon deep inelastic scattering on protons and deuterium,
and from neutrino deep inelastic scattering on nuclear targets,
which when expressed in terms of quark distributions, read
\begin{equation}
F_{2}^{\mu p}-F_{2}^{\mu n}=\frac{1}{3}x(u+\overline{u}-d-
\overline{d});
\end{equation}
\begin{equation}
F_{2}^{\mu d}=\frac{1}{2}(F_{2}^{\mu p}+F_{2}^{\mu n})=\frac{5}{18}x
(u+\overline{u}+d+\overline{d}+\frac{4}{5}s);
\label{eq:eq2}
\end{equation}
\begin{equation}
F_{2}^{\nu d}=
\frac{1}{2}(F_{2}^{\nu D}+F_{2}^{\overline{\nu} D})=x
(u+\overline{u}+d+\overline{d}+2s);
\label{eq:eq3}
\end{equation}
\begin{equation}
x F_{3}^{\nu d}=
\frac{1}{2}x(F_{3}^{\nu D}+F_{3}^{\overline{\nu} D})=x
(u-\overline{u}+d-\overline{d}),
\end{equation}
where $F_{2,3}^{\nu d}$ are converted from  $F_{2,3}^{\nu Fe}$
using a heavy-target correction factor,
with $F_{2,3}^{D}$ denoting $\frac{1}{2}(F_{2,3}^{p}
+F_{2,3}^{n})$.
These
four observables determine four combinations of parton distributions,
which can be taken to be $u+\overline{u}$, $d+\overline{d}$,
$\overline{u}+\overline{d}$ and $s$ 
by assuming $s(x)=\bar s(x)$. 
From Eqs.~(\ref{eq:eq2})
and (\ref{eq:eq3}), we
obtain the equality
\begin{equation}
\frac{5}{6}F_{2}^{\nu d}(x)-3F_{2}^{\mu d}(x)=xs(x).
\label{eq:eq5}
\end{equation}
The CTEQ Collaboration also plotted the quantity on the left-hand
side of this equation at $Q^{2}=5$ GeV$^{2}$ using data from
NMC and CCFR, as shown in Fig.~1. We thus know that the large strange
quark distributions in both the CTEQ and MRS global analyses are
natural results of
the data of the muon structure function of deuterium and
the neutrino structure function 
$F_{2}^{\nu d}$ converted from $F_{2}^{\nu Fe}$.
The smaller MRS strange quark distribution compared to that
of CTEQ is due to a renormalization factor 0.94 for the
CCFR structure function data.
We will show, in the following,
that several nuclear effects could contribute to the
right-hand side of Eq.~(\ref{eq:eq5}).

We first consider the nuclear shadowing correction in the
deuteron structure function
\begin{equation}
F_{2}^{\mu d}=\frac{1}{2}(F_{2}^{\mu p}+F_{2}^{\mu n}+\Delta
F_{2}^{\mu d}),
\end{equation}
which was chosen to ignore the shadowing term
$\Delta F_{2}^{\mu d}$ and
assumed to be Eq.~(\ref{eq:eq2}) by NMC, CTEQ, and MRS.
The sign of
$\Delta F_{2}^{\mu d}$ is negative, thus it contributes
positively to the left-hand side of Eq.~(\ref{eq:eq5}). There have
been several theoretical works on the shadowing correction
in the deuteron with different predictions about
the magnitude \cite{Zol92}-\cite{MT93}.
One work \cite{Zol92} has suggested a significant
amount of shadowing in the deuteron (up to 4\% for
$x \simeq 0.01$), whereas other calculations have predicted less
dramatic effects (2\% in Ref.~\cite{BK92} and 1\% in
Ref.~\cite{MT93}). 
In Fig.~1 we present our calculated
$\frac{3}{2}\Delta F_{2}^{\mu d}$ following Ref.~\cite{Zol92},
which gives
the most large theoretical estimate among available
works. We see that this most large estimate is
of about 30\% of the strange quark distribution measured
by CCFR, and it can explain the discrepancy
between the results by CCFR dimuon measurement and by MRS.
However,
it is too small to explain the discrepancy
between the results by CCFR and by CTEQ.

Then we analyse the corrections due to isospin symmetry
breaking in the neutrino structure function
$F_{2}^{\nu d}$,
which was converted by CTEQ and MRS
from CCFR $F_{2}^{\nu Fe}$ data
using a heavy-target correction factor.
Actually the CCFR $F_{2}^{\nu Fe}$  
data were from a combination of
neutrino and antineutrino data and it should be
$F_{2}^{\nu Fe}+F_{2}^{\overline{\nu} Fe}$.
Thus
$F_{2}^{\nu d}$ should be expressed by
\begin{equation}
F_{2}^{\nu d}(x)=(Z F_{2}^{\nu p}+N F_{2}^{\nu n})/A
=\frac{1}{2}(F_{2}^{\nu p}+F_{2}^{\nu n})+
\frac{I}{2}(F_{2}^{\nu n}-F_{2}^{\nu p}),
\end{equation}
where $I=(N-Z)/A$  is the isotopic asymmetry parameter for Fe.
If we assume isospin symmetry between the proton
and the neutron, i.e.,
\begin{equation}
F_{2}^{\nu p}(x)=F_{2}^{\nu n}(x),
\end{equation}
we can express
$F_{2}^{\nu d}$
by
\begin{equation}
F_{2}^{\nu d}(x)
=\frac{1}{2}(F_{2}^{\nu p}+F_{2}^{\nu n}),
\end{equation}
which is Eq.~(\ref{eq:eq3}).
However, it has been suggested \cite{Ma} that the isospin
symmetry breaking could be an alternative source for the Gottfried
sum rule violation reported by
the New Muon Collaboration (NMC) \cite{NMC91}. This means
that there are more sea quark in neutrons than in protons,
while the u- and d- sea symmetry is still preserved.
In this case we have
\begin{equation}
F_{2}^{\nu n}(x)-F_{2}^{\nu p}(x)
=4x[O_{q}^{n}(x)-O_{q}^{p}(x)],
\end{equation}
where $O_{q}^{n,p}(x)=\overline{u}^{n,p}(x)=
\overline{d}^{n,p}(x)$ are the u- and d- sea quark
distribution in neutrons and protons.
Because
$O_{q}^{n}(x)-O_{q}^{p}(x)$ is positive,
it contributes
positively to the left-hand side of Eq.~(\ref{eq:eq5}).
From Ref.~\cite{Ma} we
know that we need
\begin{equation}
\int_{0}^{1} {\rm d} x [O_{q}^{n}(x)-O_{q}^{p}(x)]=0.084
\end{equation}
to reproduced the observed Gottfried sum $S_{G}=0.240$
reported by NMC, by neglecting
the shadowing correction in deuterium.
If we take the shadowing correction as adopted above, we need
\begin{equation}
\int_{0}^{1}{\rm d} x[O_{q}^{n}(x)-O_{q}^{p}(x)]=0.165
\label{eq:eq6}
\end{equation}
to reproduced the observed $S_{G}$. This will give a
larger correction
to the left-hand side of Eq.~(\ref{eq:eq5}).
To estimate the magnitude of
the correction, we adopted a non-Pomeron form correction
$O_{q}^{n}(x)-O_{q}^{p}(x)=a(1-x)^{b}$, which has been analyzed
\cite{MSWn}
to study the proton-induced Drell-Yan production data
of the Fermilab experiment E772 \cite{E772}
in the isospin breaking explanation for the Gottfried
sum rule violation. By choosing $b=25$,
with $a$ adjusted to satisfy
Eq.~(\ref{eq:eq6}), we can still reproduce the E772 data
for the ratio of cross section
$R=\sigma_{\rm W}/\sigma_{\rm IS}$ and for the shape
of the differential cross section $m^{3}{\rm d}^{2}\sigma/{\rm d}
x_{\rm F}
{\rm d}m$ for
$^{2}$H. In Fig.~1 we present the calculated
$\frac{I}{2}(F_{2}^{\nu n}-F_{2}^{\nu p})$ due to isospin
symmetry breaking. We find that the correction is
of about 4\% of the strange quark distribution measured
by CCFR. It is too small to explain the discrepancy
between the results by CCFR dimuon measurement and
by CTEQ and MRS global analyses. 

We now check the
heavy-target correction factor $A(x)$ which is used to convert
$F_{2}^{\nu Fe}(x)$ to
$F_{2}^{\nu d}(x)$; i.e.,
\begin{equation}
F_{2}^{\nu d}(x)=A(x)
F_{2}^{\nu Fe}(x).
\end{equation}
Both CTEQ and MRS have adopted the correction factor $A(x)$ by
taking into
account nuclear effects
based on the muon iron/deuterium structure
function ratios observed by  EMC, NMC {\it et al.}. The exact correction
factor $A(x)$ used by MRS is presented in Tab.~I of Ref.~\cite{MRS93},
with a further normalization factor 0.94.
$A(x)$ used
by CTEQ is expressed by
\begin{equation}
A(x)=1/R(x)=[1.118 - 0.4199x - 0.3597 {\rm exp}(-22.88x) +
      1.872x^{11.27}]^{-1},
\end{equation}
where $R(x)$ is a
parametrization of
the NMC measurement of ${\rm Ca/D}_2$ and the SLAC result for ${\rm
Fe/D}_2$.
We see that both the nuclear shadowing effect and the
EMC effect have been considered in $A(x)$. 
We indicate here that an assumption of
the absence of shadowing in neutrino scattering
will reduce $F_{2}^{\nu d}$ by an amount of
$(A(x)-1)F_{2}^{\nu Fe}$
at small $x$. In Fig.~1 we present the two points of the data
$\frac{5}{6}(A(x)-1)F_{2}^{\nu Fe}(x)$. We see that the magnitude
of the two points are comparable with the difference of the
strange quark
distribution parametrized by CTEQ and that by CCFR dimuon
measurement.
However, the shadowing is expected also to occur in neutrino
scattering and there have been available WA59 data
indicating a shadowing compatible with   
predictions \cite{All89}.
The renormalization factor
0.94 in the MRS analysis is large and important
at small $x$, as can be seen Fig.~1. This is the reason for the large
difference between the CTEQ and MRS results of strange quark
distributions.  However, this normalization correction seems to be
suitable at small $x$, but too large at larger $x$.  

One conclusion from our work is that it is of essential
importance to consider all possible corrections due to
nuclear effects in the available experimental data, for good and
reliable global analyses of quark distributions. It has been
discussed in this paper two possible contributions 
to the discrepancy between
the strange quark distributions obtained by two global
analyses and the strange quark distribution measured
by the CCFR Collaboration from dimuon events in neutrino
scattering. The
shadowing in the deuteron and
the isospin symmetry breaking in the sea between the
neutron and proton could provide some corrections, but are still
too small to explain the discrepancy.  
This suggests that new mechanism is needed to solve the conflict
between the strange quark distributions of the global
analyses and those
by the CCFR Collaboration from dimuon events in neutrino
scattering. An attempt to explain this conflict has been proposed
recently as due to the strangeness quark and antiquark asymmetry
in the nucleon sea and will  
be given elsewhere \cite{Bro96}.

\vspace{2mm}

The author would like to acknowledge the valuable discussions with
J.~Botts, S.~J.~Brodsky, and A.~Sch\"afer. This work was support by
the Alexander von Humboldt Foundation and by the National
Natural Science Foundation of China under grant No.~19445004.

\break
\noindent
{\large \bf Figure Captions}
\renewcommand{\theenumi}{\ Fig.~\arabic{enumi}}
\begin{enumerate}
\item
The results of strange quark distribution
$x s(x)$ as a function of the Bjorken scaling variable $x$.
The $\bullet$ points are the CCFR data from dimuon events in
neutrino scattering for $Q^{2} \approx 5$ GeV$^{2}$.
The $\circ$ points are the CTEQ ``data'' of
$\frac{5}{6}F_{2}^{\nu d}(x)($CCFR$)-3F_{2}^{\mu d}(x)($NMC$)$.
The thick and thin solid curves are the CTEQ and MRS
parametrizations of $x s(x)$ for $Q^{2}=5$ GeV$^{2}$. 
The dashed and dotted curves
are corrections
due to the large shadowing effect in $F_{2}^{\mu d}(x)$
and the isospin symmetry breaking in $F_{2}^{\nu d}(x)$.
The $\oplus$  points are corrections from assuming 
the absence of shadowing
effect in neutrino scattering.
The dash-dotted curve is the correction due the the normalization
factor 0.94 for the MRS global analysis.
\end{enumerate}
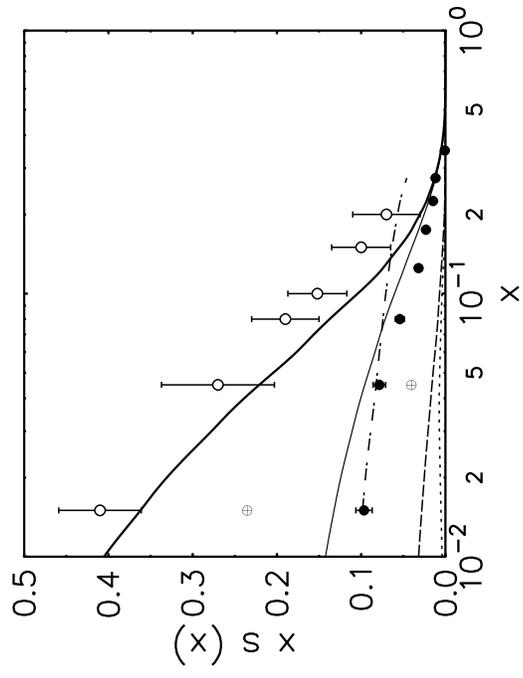
\end{document}